\documentclass[a4paper]{jpconf}
\usepackage{graphicx}
\begin{document}
\title{Search for electron liquids with non-Abelian quasiparticles}

\author{Arkadiusz W\'ojs}

\address{
    TCM Group, Cavendish Laboratory, University of Cambridge,\\
    Cambridge CB3 0HE, United Kingdom; and}

\address{
    Institute of Physics, Wroclaw University of Technology,\\
    Wyb.\ Wyspia\'nskiego 27, 50-370 Wroclaw, Poland}

\ead{arkadiusz.wojs@pwr.wroc.pl}

\begin{abstract}
We use exact numerical diagonalization in the search of fractional 
quantum Hall states with non-Abelian quasiparticle statistics.
For the (most promising) states in a partially filled second Landau 
level, the search is narrowed to the range of filling factors $7/3
<\nu_e<8/3$.
In this range, the analysis of energy spectra and correlation functions, 
calculated including finite width and Landau level mixing, supports the 
prominent non-Abelian candidates at $\nu_e=5/2$ (paired Moore--Read 
``pfafian'' state) and $12/5$ (clustered Read--Rezayi ``parafermion''
state).
Outside of this range, the noninteracting composite fermion model
with four attached flux quanta is validated, yielding the family of 
quantum liquids with fractional, but Abelian statistics.
The borderline $\nu_e=7/3$ state is shown to be adiabatically connected 
to the Laughlin liquid, but its short-range correlations are significantly
different.
\end{abstract}

\section{Introduction}

The incompressible quantum liquids (IQLs) \cite{Laughlin83} 
continue to be the subject of extensive studies ever since
the famous discovery of the fractional quantum Hall (FQH) 
effect \cite{Tsui82}.
The IQLs are formed by two-dimensional electrons placed in a 
high magnetic field $B$ which causes them to fill a particular
fraction $\nu$ of one of the lowest Landau levels (LL$_n$, 
$n=0$, 1, \dots).
The most recent storm of interest in the IQLs is motivated 
by the concept of ``topological quantum computation'' 
\cite{Kitaev03,Nayak08} employing non-Abelian statistics of 
some of the wave functions proposed for a partially filled LL$_1$.
The main idea is that the quantum information may be encoded
in (topologically) different quantum states corresponding to 
the same spatial configuration of the non-Abelian quasiparticles 
(QPs) of a given underlying IQL.
Transition between such different states would only be possible
by a global transformation of the QP braiding, thereby making 
quantum information inherently protected from decoherence caused 
by any local processes (e.g., coupling to phonons or atomic spins).
The best known candidate for a non-Abelian wave function is the
``pfaffian'' state proposed by Moore and Read \cite{Moore91,Greiter91} 
and believed to describe the FQH state in a half-filled LL$_1$.
Other wave functions with different complexities of braiding statistics 
have also been proposed \cite{Read99,Moller08a,Bonderson08,Bonderson09}.
The convincing demonstration of non-Abelian statistics in a real 
physical system has clearly risen to the challenge of greatest 
importance.

The reason to search for the signatures of non-Abelian statistics 
in LL$_1$ is that, on the one hand, the partially filled LL$_0$ 
are successfully described by the composite fermion (CF) theory 
\cite{Jain89} predicting fractional but Abelian QPs and, on the 
other hand, that higher LLs favor ordered electron phases.
Crucial recent experiments in LL$_1$ include confirmation \cite{Dolev08} 
of the anticipated QP charge of $e/4$ for the half-filled 
state at $\nu_e=5/2$ and careful measurements \cite{Pan08,Dean08} 
of the minute excitation gaps.
In theory, some of the most recent advances are related to the role 
of layer width \cite{Peterson08} and LL mixing \cite{LLmix,Simon08} 
in real systems, and the nature of the QPs \cite{Toke07,Bernevig08}.

It is quite remarkable that, despite intensive studies, connection 
of the FQH states in LL$_1$ ($\nu_e=5/2$, $7/3$, 
$12/5$, or $11/5$) to the few proposed wave functions 
remains yet to be conclusively established.
In fact, in some cases it seems only tentatively assumed for the 
lack of other candidates.
This is an urgent problem, as the anticipation of non-Abelian 
statistics in nature is largely driven by the connection of some 
of these wave functions to the particular conformal field theories.
A wealth of IQLs found in various systems (electrons or CFs at 
different fillings of different LLs, in layers of varied width $w$) 
also invites a more general question of possible IQLs with arbitrary 
LL filling $\nu$ and interaction $V$.

This paper is an extension of our brief communication \cite{trans}.
We report here on the use of large scale exact diagonalization in
the search of IQLs with non-Abelian statistics.
We demonstrate that non-Abelian IQLs in LL$_1$ can only emerge 
in the narrow, particle-hole symmetric range of filling factors 
$7/3<\nu_e<8/3$.
In this range, the known non-Abelian candidates at $\nu_e=5/2$ 
and $12/5$ are closely examined (including such previously
neglected realistic effects as finite layer width and LL mixing)
and found to have favorable correlation energies.
Outside of this range, the LL$_1$ hosts the family of Abelian 
ground states of noninteracting CFs each carrying four magnetic 
flux quanta, repeating the states known from the lowest LL.
The borderline $\nu=1/3$ ground state in LL$_1$ is adiabatically 
connected to the Laughlin state of LL$_0$, but it has a smaller gap
and distinct short-range correlations.

\section{Model}

We consider the systems of $N$ spin-polarized fermions (electrons) 
on a Haldane sphere with unit radius and the magnetic monopole of 
strength $2Q(hc/e)$ inside \cite{Haldane83}.
In this geometry, LL$_n$ is a shell of angular momentum $\ell=Q+n$, 
and different $N$-body wave functions at the same $\nu$ are (unlike
on a torus) conveniently distinguished by a `shift' $\gamma$ between 
the LL degeneracy and $\nu^{-1}N$ (i.e., $2l=\nu^{-1}N-\gamma$).
In contrast to the previous exact diagonalization studies we begin 
by searching the universality classes $(\nu,\gamma)$ of the gapped 
ground states with arbitrary interactions $V$ rather than confining 
ourselves to the particular physical systems defined by $n$, $w$, etc.

It is a trivial fact that the many-body dynamics in a degenerate 
LL is completely determined by an interaction pseudopotential, 
defined \cite{Haldane83} as the dependence of the pair energy $V$ 
on the relative angular momentum $m=1$, 3, \dots.
Less obviously, $V_m$ induces particular correlations only through 
its deviation from a reference ``harmonic pseudopotential'' given
by a straight line over the consecutive $m$'s \cite{five-half}.
Hence, the low-energy spectra of $V_m$ can be accurately reproduced 
by a suitable effective pseudopotential $U_m$ with only a few 
appropriate coefficients.

In our calculations we have used $U=[U_1,U_3,U_5]$.
Higher-order terms have been ignored, as they are essentially 
irrelevant for the dynamics in a liquid phase with short-range 
correlations.
On the other hand, going beyond the (earlier used) $U_1$ and $U_3$ 
was needed for an improved description of the two lowest (electron 
or CF) LLs known to host IQL states.
At the same time, it still allowed for useful graphical representation 
of the ground state properties in an effectively two-dimensional space 
of the (normalized, $\sum_m U_m=1$) parameters $U_m$.

\section{Maps of the gap for arbitrary interaction}

\begin{figure}[t]
\centering
\includegraphics[width=24pc]{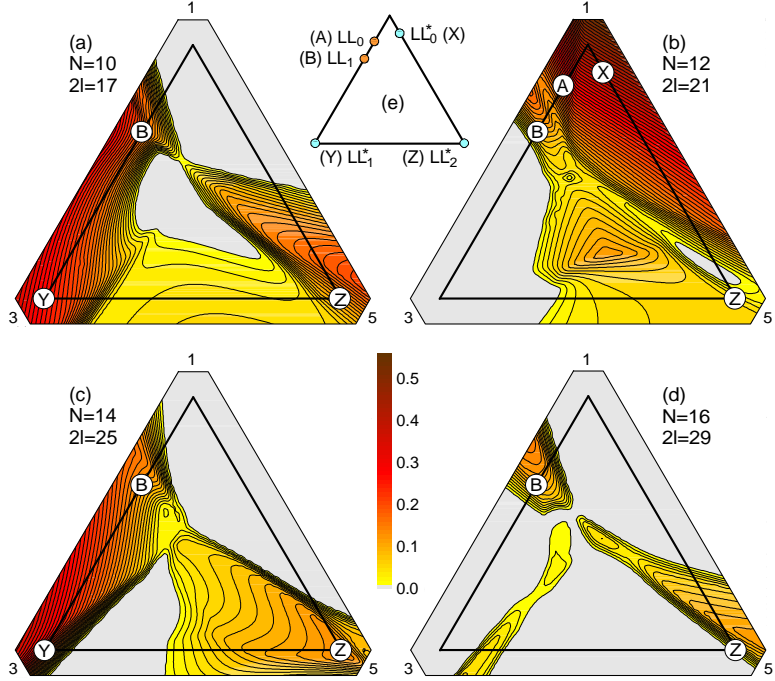}
\caption{\label{fig1}
Ternary contour plots of the neutral excitation gap $\Delta$
for $N=10$, 12, 14, 16 fermions in a half-filled Landau 
level ($\nu=1/2$) with shell angular momentum $\ell$ 
corresponding to the shift $\gamma=3$ of the Moore--Read 
``pfaffian'' wave function, interacting by model pseudopotential 
$U_m$.
In each plot, three corners of the big triangles correspond to 
$U_m=\delta_{m,\mu}$ with $\mu=1$, 3, 5 marked in each corner.
Points relevant for the actual interactions in different electron 
or CF Landau levels are marked on a triangle in frame (e).
Different candidate incompressible states are indicated.}
\end{figure}

Searching for the series of gapped ground states with particular
filling factor $\nu$ and shift $\gamma$, we looked at various 
finite systems $(N,2\ell)$.
A few maps of the `neutral' energy gap $\Delta$ for $\nu=1/2$ 
and $\gamma=3$ (i.e., $2\ell=2N-3$, as appropriate for the 
Moore--Read pfaffian state) are shown in Fig.~\ref{fig1}.
The gap $\Delta$ is defined as the energy difference from the ground 
state to the first excited state in the same spectrum as long as the 
ground state happens to be nondegenerate (i.e., has zero total angular 
momentum, $L=0$); otherwise $\Delta$ is set to zero.
The IQL candidates are the islands of positive $\Delta$ repeating 
regularly in the same area of the map for different values of $N$.
Their location on the map must be compared with the actual
pseudopotentials in different LLs, as indicated in frame (e).
In particular, (A) and (B) mark the appropriate positions of the 
electron pseudopotentials in LL$_0$ and LL$_1$ (the former
dominated by $V_1$; the latter roughly linear between $m=1$ and 5),
and (X), (Y), and (Z) mark the same for the CFs in their effective
shells LL$^*_0$, LL$^*_1$, and LL$^*_2$.

The appearance of a significant gap around the point (B) in all 
maps in Fig.~\ref{fig1} (i.e., for each $N$) confirms quite 
definitively the earlier expectation that the Moore--Read ground 
state forms for a class of pseudopotentials close to that of LL$_1$.
It also demonstrates that its accuracy depends sensitively on 
the fine-tuning of the leading $V_m$'s, achieved (for example) 
by adjusting the layer width $w$ \cite{Peterson08}.
Remarkably, the maps in Fig.~\ref{fig1} also preclude the Moore--Read 
state at the half-filling of other LLs (e.g., of the second CF LL, 
called LL$^*_1$, characterized by a dominant repulsion at $m=3$ 
\cite{Lee01}).
Clearly, the microscopic origin of the FQH state observed at 
$\nu_e=3/8$ \cite{Pan03} (in the CF picture, corresponding 
to $\nu=1/2$ in LL$^*_1$) must be different.

\begin{figure}[t]
\begin{minipage}{18pc}
\includegraphics[width=18pc]{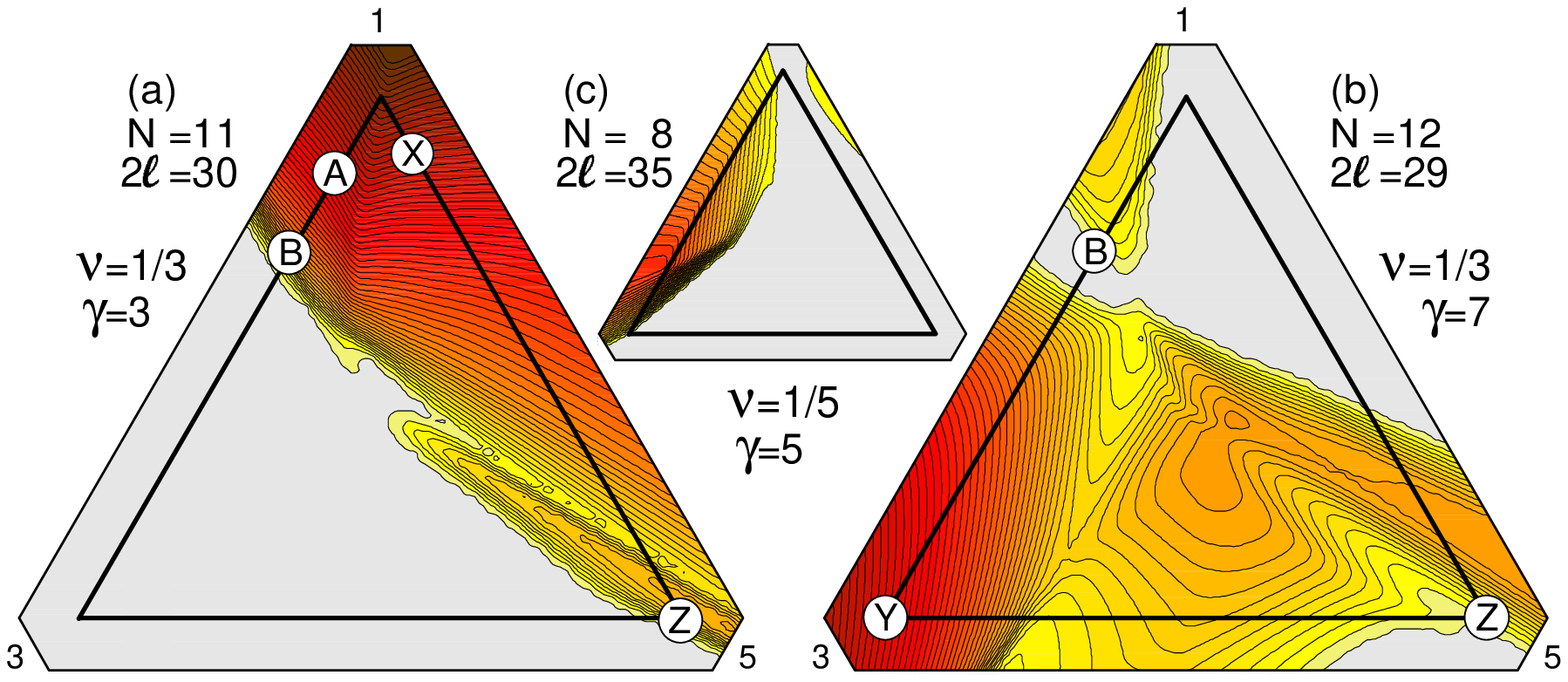}
\caption{\label{fig2}
Similar to Fig.~\ref{fig1}, but for Laughlin filling 
fractions $\nu=1/3$ (two candidate states with 
$\gamma=3$ and 7) and $\nu=1/5$.}
\end{minipage}
\hspace{2pc}
\begin{minipage}{18pc}
\includegraphics[width=18pc]{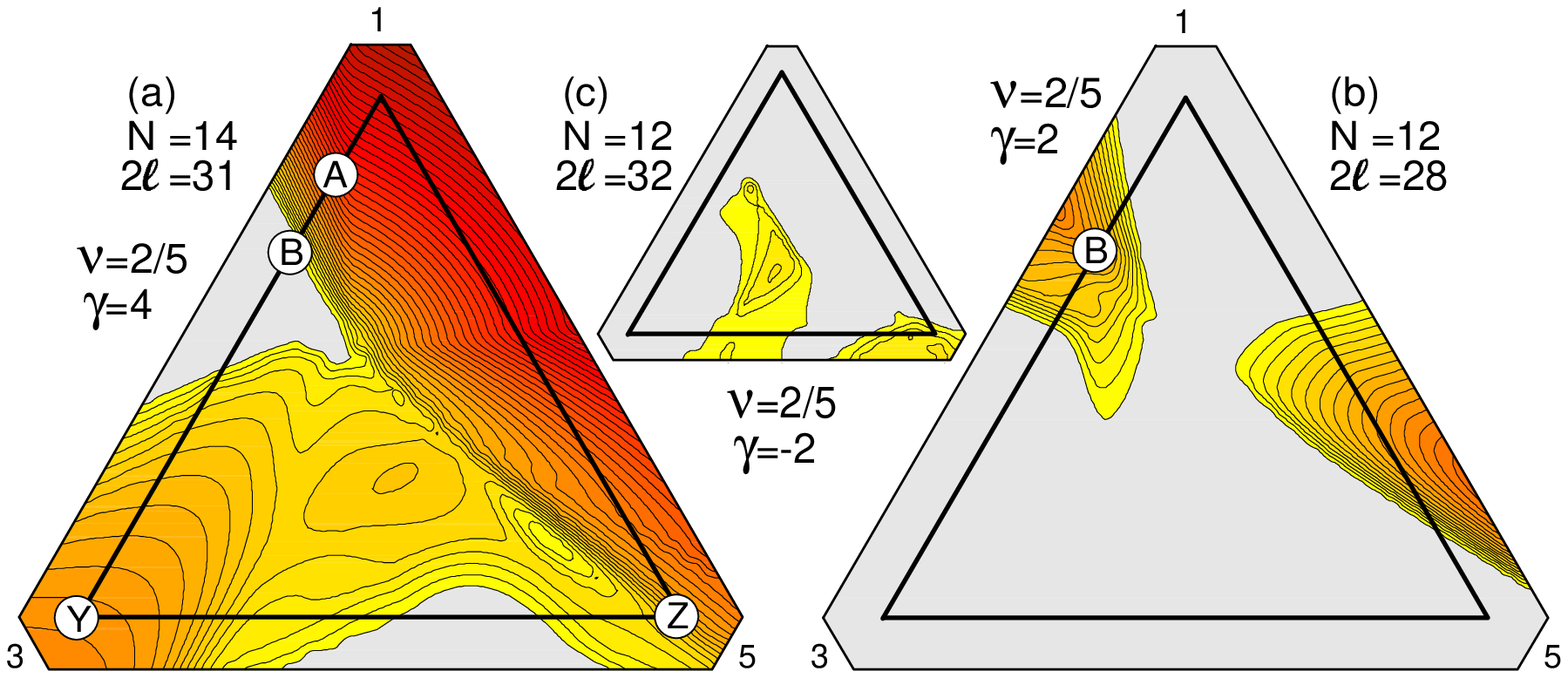}
\caption{\label{fig3}
Similar to Fig.~\ref{fig1}, but for the $\nu=2/5$ 
states with $\gamma=4$ (Jain), 2 (Bonderson--Slingerland), 
and $-2$ (Read--Rezayi).}
\end{minipage}
\end{figure}

Analogous maps of $\Delta$ for the Laughlin filling fractions
$\nu=1/3$ and $1/5$ are shown in Fig.~\ref{fig2}.
In (a), the universality class of the Laughlin wave function
($\nu=1/3$ and $\gamma=3$) is shown to cover a large
part of the map, including (A) and (X), and possibly also 
(B) and (Z).
Point (B), the most interesting for the present analysis, 
falls just inside the island of positive gap $\Delta$,
suggesting connection of the $\nu_e=7/3$ FQH state to 
the Laughlin $\nu=1/3$ liquid in LL$_1$.
Point (X) is relevant for the Laughlin $\nu=1/3$ state 
of the CF vacancies in LL$^*_0$, i.e., to the robust Jain 
state at $\nu_e=2/7$.
Remarkably, the Laughlin ground state does not occur around 
point (Y) corresponding to the $\nu=1/3$ filling of 
LL$^*_1$.
The nature of the rather fragile FQH state observed at the 
corresponding fraction $\nu_e=4/11$ \cite{Pan03} must 
therefore be different.
In (b), the $\nu=1/3$ paired (non-Laughlin) state with 
$\gamma=7$, proposed earlier for both LL$_1$ \cite{five-half} 
and LL$^*_1$ \cite{clusters}, is tested.
The gap around (Y) is quite suggestive that it may indeed 
describe the FQH state at $\nu_e=4/11$ \cite{Pan03}.
On the other hand, its relevance for LL$_1$ seems doubtful.
In (c), any positive $U=[U_1,U_3,0]$ yields an exact Laughlin 
state at $\nu=1/5$.
Our map of $\Delta$ confirms that it is true description of 
the FQH states in both LL$_0$ and LL$_1$ \cite{Ambrumenil88}.
On the other hand, its relevance to the FQH effect observed 
in LL$^*_1$ (i.e., at $\nu_e=6/17$) \cite{Pan03} is rather 
unlikely.

Finally, the maps of $\Delta$ for the $\nu=2/5$ filling 
have been shown in Fig.~\ref{fig3}.
In (a), the Jain series of noninteracting CF states with 
$\gamma=4$ correctly represents the ground state in LL$_0$ 
(and LL$^*_0$).
Point (B) corresponding to LL$_1$ lies just outside the borders 
of the island of $\Delta>0$, leaving the question of relevance 
of the non-interacting CF picture at $\nu_e=12/5$ open.
Two other candidate ground states at $\nu=2/5$ are the 
parafermion state with $\gamma=-2$ \cite{Read99} and a more 
recently proposed state with $\gamma=2$ \cite{Bonderson08}.
Especially for the latter state, frame (b) appears suggestive of 
a gap emerging around (B).
Clearly, the competition between these three candidate ground 
states in LL$_1$ is not convincingly resolved based on the 
maps of $\Delta$ alone.
On the other hand, the identification of the true ground state
is crucial, because the candidates with $\gamma=\pm2$ are both 
non-Abelian, in contrast to the $\gamma=4$ Jain state.
More careful analysis will follow in subsequent sections.

\section{Maps of the amplitudes for arbitrary interaction}

\begin{figure}[t]
\centering
\includegraphics[width=9pc]{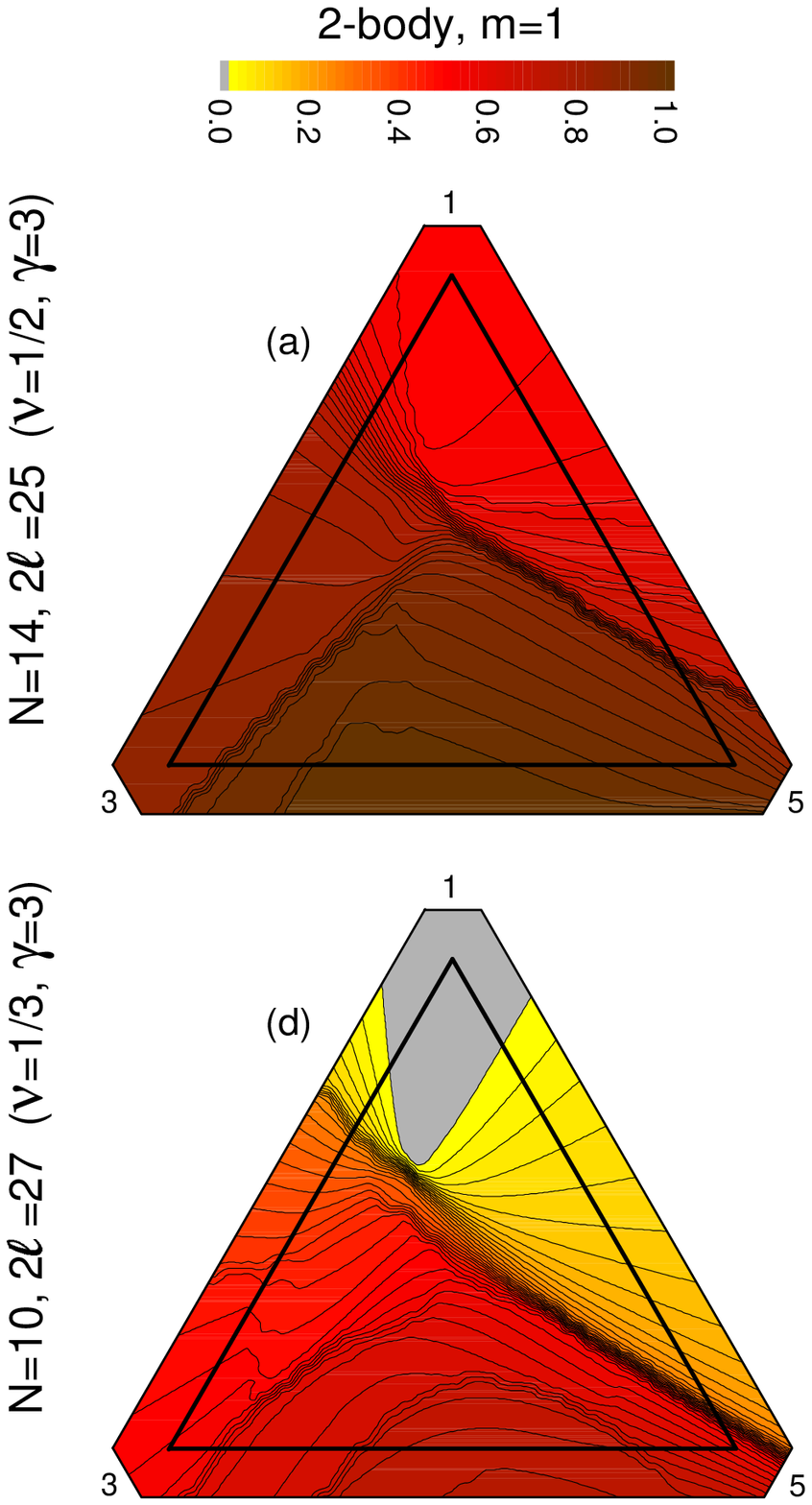}
\includegraphics[width=9pc]{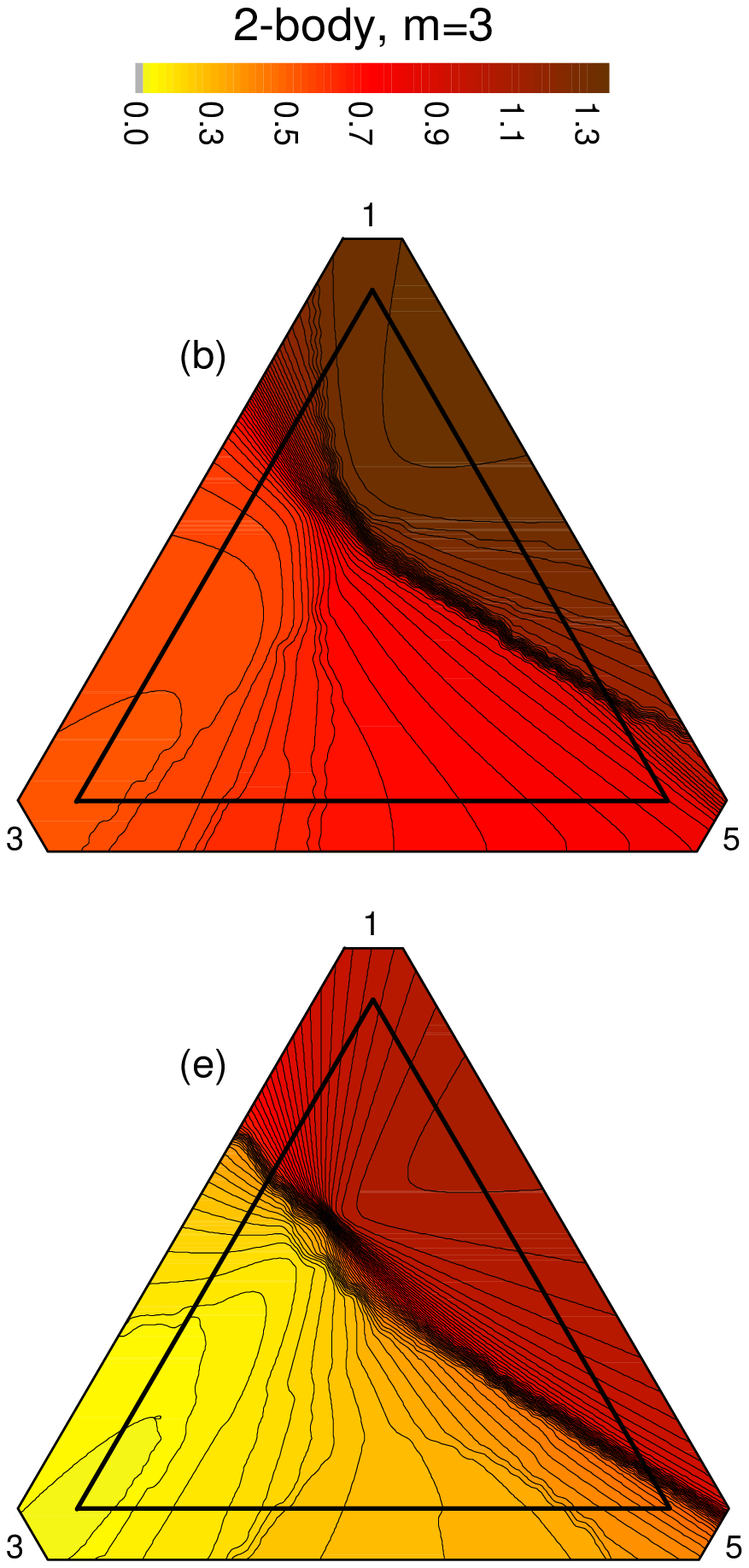}
\includegraphics[width=9pc]{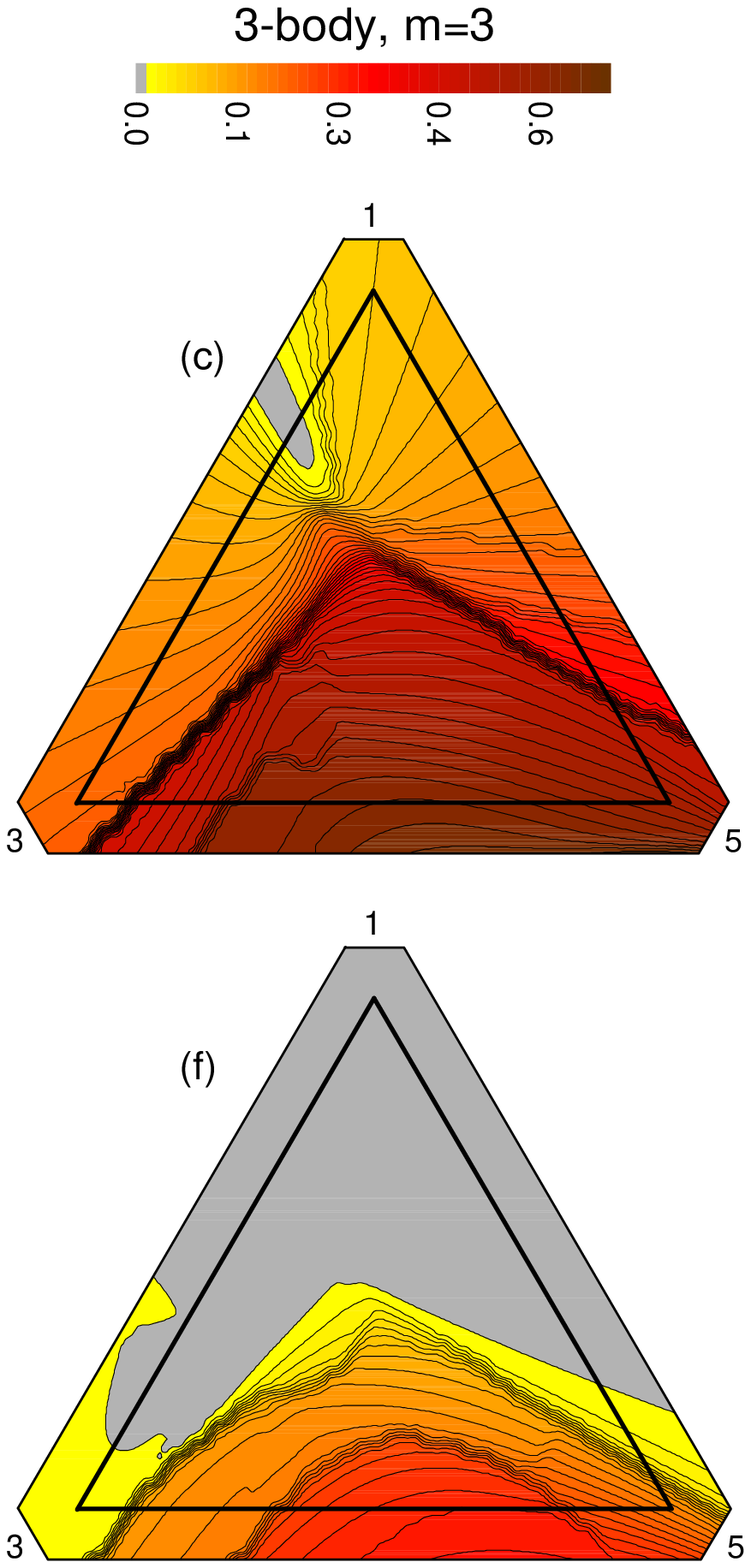}
\caption{\label{fig4}
Ternary contour plots (similar to Fig.~\ref{fig1}) of the pair and 
triplet amplitudes (labeled by the relative angular momentum $m$) 
for the universality classes of Laughlin and Moore--Read wave 
functions (filling factors $\nu=1/3$ and $1/2$, respectively; 
shift $\gamma=3$ in both cases).}
\end{figure}

Maps similar to those in Figs.~\ref{fig1}--\ref{fig2} can also
be used to show dependence of other parameters of the spectrum 
on the form of interaction.
For example, in Fig.~\ref{fig4} we plot maps of the leading pair 
and triplet Haldane amplitudes for the ground states of 
$(\nu,\gamma)=(1/2,3)$ and $(1/3,3)$.
The amplitudes are defined \cite{3body} as the fractions of the 
pairs or triplets with a given relative angular momentum $m$
(for the pairs, $m=1$, 3, 5, \dots\  is a measure or an average 
square distance; for the triplets, $m=3$, 5, 6, \dots\ measures 
an average area).
These (discrete) correlation functions are particularly useful
in identifying Laughlin and Moore--Read states, as they both are 
unique zero-energy states of simple repulsions: the former at 
the minimum pair angular momentum $m=1$, the latter at the minimum
triplet angular momentum $m=3$ (hence, their corresponding 
amplitudes vanish exactly).
The emergence and location of the islands of essentially zero 
amplitude in Fig.~\ref{fig4}(c) and (d) provides additional
and quite decisive support for the Laughlin $\nu=1/3$
and Moore--Read $\nu=1/2$ ground states (universality 
classes) in LL$_0$ and LL$_1$, respetively. 
On the other hand, other questions, such as of a Laughlin state 
in LL$_1$, remain open until the competition with other possible 
states (with the same $\nu$ but different $\gamma$) can be 
resolved.

\section{Analysis of correlation energies in LL$_0$ and LL$_1$}

\begin{figure}[t]
\centering
\includegraphics[width=24pc]{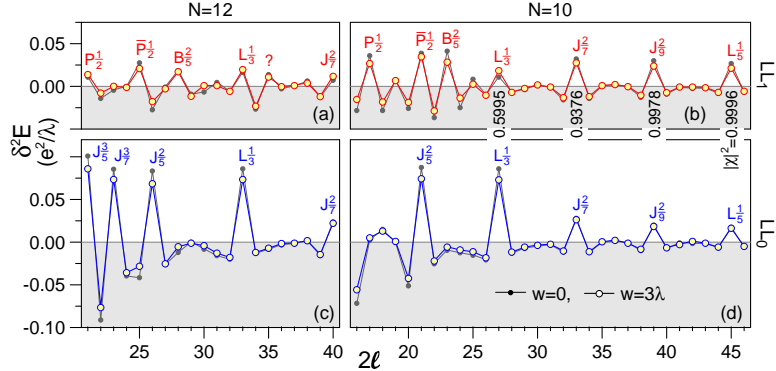}
\caption{\label{fig5}
Symmetric second-order differences $\delta^2E$ of the ground-state 
energy per particle $E$ for $N=10$ and 12 electrons in the lowest 
and second Landau level (LL$_0$ and LL$_1$), as a function of shell 
angular momentum $\ell$, for layer widths $w=0$ and $3\lambda$.
Candidate incompressible states are marked as `$\Phi\nu$', 
where $\Phi={\rm P}$, $\overline{\rm P}$, L, J, B denote 
Moore--Read pfaffian, anti-pfaffian, Laughlin, Jain, and
Bonderson--Slingerland states.
Squared overlaps $|\chi|^2$ between the states repeating in both 
LLs at $\nu\le1/3$ are indicated.
$\lambda$ is the magnetic length.}
\end{figure}

Guided by the maps of gaps and amplitudes we now move our focus 
to the FQH states in a partially filled LL$_1$.
In Fig.~\ref{fig5} we seek confirmation of the IQL candidates in 
the downward cusps of the dependence of the ground-state correlation 
energy per particle $E$ on the LL degeneracy (for a fixed number 
of electrons $N$).
The correlation energy $E$ is calculated from the total Coulomb 
energy $\mathcal{E}$ of $N$ electrons in a LL shell with a given 
$\ell$ by adding the energy of attraction to the uniform 
charge-compensating background and dividing by $N$.
In the calculation for finite width $w$ of a quasi-2D electron layer, 
the Coulomb matrix elements were computed assuming infinite-well 
confinement in the perpendicular direction, i.e., for the 
charge-density profile of $\varrho(z)=(2/w)\cos^2(\pi z/w)$.
The cusps in $E$ are most pronounced in the plots of a symmetric 
difference $\delta^2E_{2\ell}=E_{2\ell-1}+E_{2\ell+1}-2E_{2\ell}$.
For an IQL, its (positive) value gives the QP gap $\tilde\Delta$ 
(energy needed to create a pair of noninteracting QPs of total 
charge zero) times the number of QPs created per flux quantum.

Evidently, Fig.~\ref{fig5} complements the maps of 
Fig.~\ref{fig1}-\ref{fig4} in the identification of the universality 
classes of particular IQLs.
For example, it shows peaks in $\delta^2E_{2\ell}$ which signal the
Laughlin $\nu=1/3$ and Moore--Read $\nu=1/2$ IQLs in 
LL$_1$.
But Fig.~\ref{fig5} also does more, by revealing the following 
connection between quantum statistics and the filling factor in 
a partially filled LL$_1$.
At $\nu<1/3$ the same Laughlin and Jain IQLs of filled shells
of noninteracting CFs with four flux quanta occur in LL$_1$ and LL$_0$.
The corresponding states in LL$_1$ and LL$_0$ have high overlaps 
(calculated by replacing the electron positions by the guiding centers)
and similar QP gaps $\tilde\Delta$.
This similarity, earlier pointed out in Ref.~\cite{Ambrumenil88}, 
is caused by a sufficiently high pseudopotential coefficient $V_1$ 
(forcing the avoidance of the $m=1$ pair state) and a similar behavior 
of the pseudopotential al long range, $V_{m\ge3}$, in the two lowest 
LLs.
Importantly, this similarity validates the noninteracting CF model 
\cite{Jain89} with four flux quanta attached to each electron in 
LL$_1$ (in addition to LL$_0$ where its accuracy is well-known).

In contrast, at $\nu>1/3$ the Jain sequence of noninteracting 
CF states in LL$_0$ is replaced in LL$_1$ by a different set of IQLs, 
including the Moore--Read pfaffian at $\nu=1/2$, the anti-pfaffian 
(pfaffian's particle-hole conjugate) at the same $\nu=1/2$ but with 
a different shift $\gamma=-1$, and a Bonderson--Slingerland $\nu=2/5$ 
state with $\gamma=2$.
The breakdown of the noninteracting two-flux CF model in LL$_1$
opens, exclusively at $1/3<\nu<2/3$, possibility for other IQLs, 
including several suggested more exotic states with various 
non-Abelian QP statistics.
The borderline $\nu=1/3$ state, separating the Abelian from 
(possibly) non-Abelian states, has a moderate overlap with the 
Laughlin state of LL$_0$, despite falling into the same class 
of $\gamma=3$.

\begin{figure}[t]
\centering
\includegraphics[width=24pc]{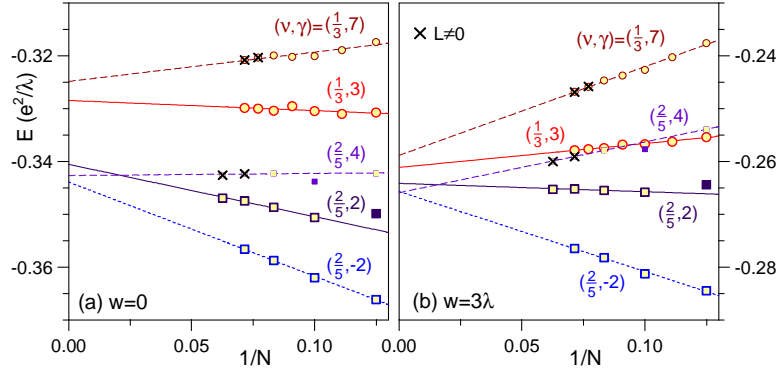}
\caption{\label{fig6}
Size extrapolation of the ground-state correlation energies 
per particle $E$ calculated for $N$ electrons in the second 
Landau level (LL$_1$), for the layer widths $w=0$ (a) and 
$3\lambda$ (b).
Competing series of candidate incompressible ground states 
with filling factors $\nu=1/3$ and $2/5$ are 
distinguished by their shifts $\gamma$.}
\end{figure}

Having narrowed the search for non-Abelian IQLs in LL$_1$ to the 
range of $1/3<\nu<2/3$, let us now try to resolve more decisively 
the competition between different candidate wave functions at 
filling factors $\nu=1/3$ and $\nu=2/5$ (at $\nu=1/2$, the 
Moore--Read pfaffian appears to have no competition).
In Fig.~\ref{fig6} we attempt to extrapolate to an infinite system size
(to $N^{-1}\rightarrow0$, i.e., to the planar geometry) the ground-state 
correlation energies per particle $E$.
To improve convergence, the plotted energies $E$ have been rescaled by 
$\sqrt{2Q\nu/N}$ so as to ensure equal units $e^2/\lambda$ for each 
$\gamma$; here $\lambda=\sqrt{hc/eB}$ is the magnetic length).
In the extrapolation for $\nu=2/5$ we only used the open squares, 
discarding one state aliased with the anti-pfaffian (a finite system 
defined by $N$ and $2\ell$ may sometimes represent states with different 
combinations of $\nu$ and $\gamma$) and one apparently suffering from 
the small size.
When choosing the IQL ground state we also checked if a given candidate 
series $(\nu,\gamma)$ consistently has a nondegenerate (i.e., one with 
$L=0$) ground state.
In those cases when the particular $(N,2\ell)$ ground state had $L\ne0$,
we crossed out the corresponding data point in Fig.~\ref{fig6} to mark
that the series containing it is unlikely to describe an IQL in the
thermodynamic limit.

For $\nu=1/3$ Fig.~\ref{fig6} eliminates the (paired) $\gamma=7$
series as a viable candidate in LL$_1$ (it remains plausible in LL$^*_1$ 
which, however, will not be further discussed here), leaving the 
(universality class of the) Laughlin state as the only acceptable 
option.
For $\nu=2/5$, the Jain series appears to extrapolate to a 
competitive energy, but it no longer has $L=0$ for the larger $N$,
and hence should probably be discarded.
The remaining two candidates both consistently have $L=0$ and both
extrapolate to nearly the same $E$ for large $N$ (the difference in
favor of the Read--Rezayi state with $\gamma=-2$ is comparable to
the error of extrapolation).
In realistic conditions this near degeneracy is likely to be removed 
by their different susceptibilities to the LL mixing.
To explore this idea, we estimated the appropriate energy correction 
$dE$ by including in the diagonalization additional states involving 
a single cyclotron excitation \cite{LLmix}.
Specifically, in addition to the configurations with completely filled 
LL$_0$ (with both spin-$\uparrow$ and spin-$\downarrow$), $N$ polarized 
electrons in LL$_1$, and empty higher LLs, we have also included 
configurations with one electron promoted from either LL$_0$ (with
either spin) or from LL$_1$ to the next higher LL.
It should be kept in mind that inclusion of only a single cyclotron 
excitation is not a rigorous treatment of the LL mixing.
However, it is expected to reveal a possible difference in the 
susceptibility of the competing states to this process.
In the calculation, we have assumed the Coulomb-to-cyclotron energy 
ratio of $\beta\equiv(e^2/\lambda)/(\hbar\omega_c)=1.56$, corresponding 
to $B=2.6$~T.
Due to a large size of the Hilbert space, the values of $dE$ could 
only be calculated for $N\le10$.
However, we found that the corrections $dE$ are far less size-dependent 
than the base energies $E$. 
It is therefore justified to apply the $N=10$ estimates of $dE$
to the values of $E$ extrapolated from $N\le16$.

\begin{table}[b]
\caption{\label{tab1}
   Extrapolated correlation energies per particle $E$ 
   (in the units of $e^2/\lambda$) in several series 
   of candidate ground states (distinguished by shifts 
   $\gamma$) at filling factors $\nu=1/3$ and 
   $2/5$ in LL$_1$, for layer widths $w=0$ and 
   $3\lambda$, calculated without LL mixing.
   Also, corrections $dE$ due to LL mixing (at $B=2.6$~T), 
   estimated for $N=10$ (except for $N=8$ for $\nu=2/5$ 
   and $\gamma=4$).}
\centering
\begin{tabular}{rrrrrrrr}
\br
$(\nu,\gamma)$ & \rule{2mm}{0mm} & 
$(1/3,3)$ & 
$(1/3,7)$ & \rule{5mm}{0mm} & 
$(2/5,-2)$ &
$(2/5,2)$ &
$(2/5,4)$ \\
\mr
$w=0$: \\
$E$    & & $-0.3285$ & $-0.3249$ & & $-0.3439$ & $-0.3406$ & $-0.3427$ \\
$dE$   & & $-0.0268$ & $-0.0218$ & & $-0.0294$ & $-0.0239$ & $-0.0227$ \\
$E+dE$ & & $-0.3553$ & $-0.3467$ & & $-0.3733$ & $-0.3645$ & $-0.3654$ \\
\mr
$w=3\lambda$: \\
$E$    & & $-0.2611$ & $-0.2589$ & & $-0.2657$ & $-0.2642$ & $-0.2659$ \\
$dE$   & & $-0.0087$ & $-0.0077$ & & $-0.0095$ & $-0.0085$ & $-0.0071$ \\
$E+dE$ & & $-0.2698$ & $-0.2666$ & & $-0.2752$ & $-0.2727$ & $-0.2730$ \\
\br
\end{tabular}
\end{table}

The results are presented in Tab.~\ref{tab1}.
Clearly, in addition to having the lowest $E$, the $\gamma=-2$ parafermion 
state also has the largest $|dE|$, and therefore it is quite convincingly 
predicted to define the universality class of the $\nu=2/5$ ground 
state in LL$_1$.
This is of considerable importance because this state is the only known 
candidate IQL \cite{Nayak08} whose braiding rules are sufficiently 
complex to allow quantum computation (unlike, e.g., the Moore--Read state).

\begin{figure}[t]
\centering
\includegraphics[width=24pc]{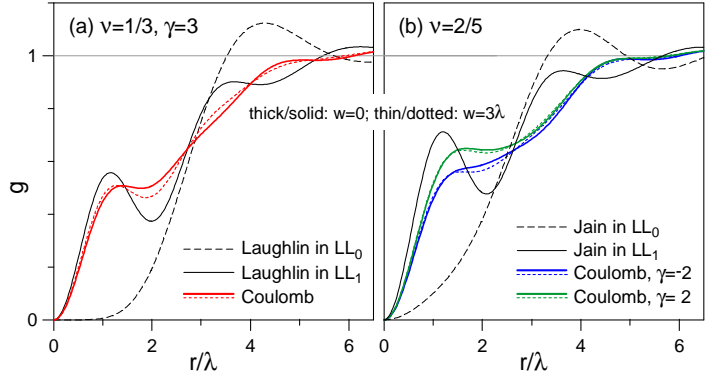}
\caption{\label{fig7}
Pair-correlation functions of the candidate incompressible
ground states at $\nu=1/3$ (a) and $2/5$ (b) 
in the second Landau level (LL$_1$).
Coulomb ground states are compared with the exact 
Laughlin and approximate Jain states in LL$_0$ and LL$_1$
(constructed as the ground states of a model interaction 
pseudopotential with pair repulsion only at $m=1$).}
\end{figure}

\begin{figure}[t]
\centering
\includegraphics[width=24pc]{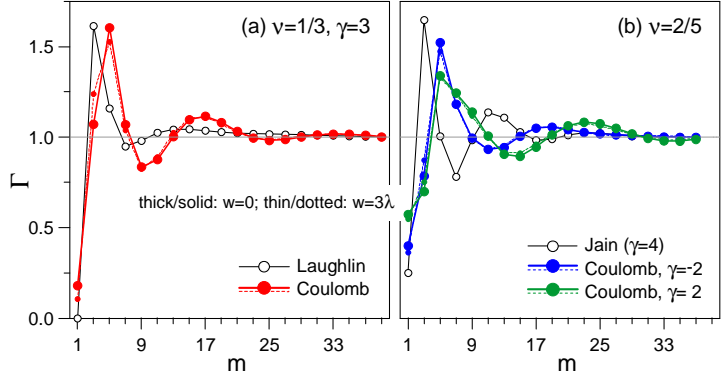}
\caption{\label{fig8}
Renormalized pair amplitudes $\Gamma$ of the same candidate 
states in LL$_1$ as in Fig.~\ref{fig7}.}
\end{figure}

Let us now look at the pair correlation functions of the established
IQL ground states at $\nu=1/3$ and $2/5$ in LL$_1$.
The results calculated for the largest available systems ($N=14$ or
16, depending on a given state, as is clear from Fig.~\ref{fig6})
are shown in Fig.~\ref{fig7}.
At $\nu=1/3$, despite being adiabalitally connected to the
Laughlin state (i.e., having the same $\gamma=3$), the Coulomb 
ground state has significantly different short-range correlations.
Specifically, the zigzag in $g(r)$ at $1<r/\lambda<2$ present in 
the exact Coulomb state in LL$_1$, caused by the particular form 
of the single particle wave functions and showing also in $g(r)$ 
for the full LL$_1$, gives way to a plateau in the case of the 
actual Coulomb ground state.
(We should also clarify here that the Laughlin state in any LL is
defined in a standard way as the zero energy state of the $m=1$ 
pair repulsion, and its pair correlation function is drawn using
the appropriate single-particle wave functions, e.g., of LL$_0$ 
or LL$_1$.)
This agrees with only moderate squared overlaps of the 
$\nu=1/3$ ground state in LL$_1$ with the Laughlin wave 
function, pointed out ourlier:
0.292 (0.501), 0.253 (0.510), 0.333 (0.549) for $N=10$, 12, 14, 
for $w=0$ ($3\lambda$).
The magneto-roton band is also absent in the spectra of LL$_1$,
and the low-energy states resembling Laughlin QEs and QHs are 
found at $2\ell=3N-3\mp1$, but they are not generally the lowest 
states in their spectra. 
A similar difference (removal of the characteristic zigzag) between 
the Jain state (approximated as the ground state of the $m=1$ pair 
repulsion for $\gamma=4$) and the competing Coulomb ground states 
with $\gamma=-2$ and 2 is also found at $\nu=2/5$ in LL$_1$.
At both filling factors dependence on $w$ is insignificant.

Difference in short-range correlations between Laughlin and Jain 
states of LL$_1$ on one hand, and the $\nu=1/3$ and $2/5$ Coulomb
ground states of LL$_1$ on the other, was also earlier apparent 
in the leading amplitudes in Fig.~\ref{fig4}(d,e).
It is evident in the comparison of renormalized pair amplitudes 
$\Gamma_m$ calculated for slightly larger systems (same as in 
Fig.~\ref{fig7}), which have been shown in Fig.~\ref{fig8}.
Here, the coefficients $\Gamma_m$ have been converted from the 
usual Haldane amplitudes on a sphere $G_m$ to their planar
counterparts, as described in Ref.~\cite{3body}.
Specifically, $\Gamma_m=1+(G_m-G^{\rm full}_m)/G^{\rm full}_1$,
where $G^{\rm full}_m=(4\ell+1-2m)/\ell/(2\ell+1)$ describes
a full LL.

\section{Conclusion}

We have carried out extensive exact diagonalization studies
(including finite layer width and LL mixing) of the fractional 
quantum Hall states in a partially filled second Landau level 
(LL$_1$), searching for non-Abelian incompressible quantum 
liquids (IQLs).
We have found the range of filling factors $1/3<\nu<2/3$ in LL$_1$ 
in which the emergence of non-Abelian statistics is possible.
Inside this range, we have demonstrated that the spin-polarized 
ground states at $\nu=1/2$ and $2/5$ are described 
by the non-Abelian ``pfaffian'' and ``parafermion'' wave 
functions.
Outside of it, the Jain states of noninteracting CFs with four
attached magnetic flux quanta repeat in both lowest LLs, thus 
precluding more exotic phases.
The borderline $\nu=1/3$ state is adiabatically connected 
to the Laughlin liquid but has a smaller gap and distinct 
short-range correlations.

\section*{Acknowledgment}

The author thanks G. M\"oller, N. Cooper, S. Simon, A. Stern, 
and G. Gervais for many insightful comments, and acknowledges 
support from EU under the Marie Curie Intra-European Grant 
No.\ PIEF-GA-2008-221701 and from the Polish MNiSW under grant 
N-N202-1336-33.

\end{document}